\documentclass[12pt]{article}
\usepackage{xcolor}
\usepackage{scicite}
\usepackage{times}
\usepackage{graphicx}
\usepackage{amsmath}
\usepackage{subcaption}
\newenvironment{sciabstract}{%
\begin{quote} \bf}
{\end{quote}}

\newcounter{lastnote}

\title{Megabar Shocks on a Benchtop}

\author
{Yun Kai$^{1,2,}$\textsuperscript{\textdagger}, Min Kim $^{6}$, Senpeng Lin$^{1,2}$, Zhen Sang$^{1,2}$, Jacob Gordon $^{1,5}$, \\
Steven E. Kooi$^{2}$, Alexei Maznev$^{1}$, Thomas Pezeril$^{4}$, Cyril Williams $^{2, 3}$, \\
Tom Lockard $^{6}$, Damian Swift $^{6}$, Keith A. Nelson$^{1, 2\ast}$\\
\\
\normalsize{$^{1}$Department of Chemistry, Massachusetts Institute of Technology, Cambridge, MA 02139, USA,}\\
\normalsize{$^{2}$Institute for Soldier Nanotechnologies, Massachusetts Institute of Technology,}\\
\normalsize{Cambridge, MA 02139, USA}\\
\normalsize{$^{3}$DEVCOM Army Research Laboratory, Aberdeen Proving Ground, MD 21005, USA}\\
\normalsize{$^{4}$Institut de Physique de Rennes, UMR CNRS 6251, Université de Rennes, 35042 Rennes, France}\\
\normalsize{$^{5}$University of Massachusetts Amherst, MA 01003, USA}\\
\normalsize{$^{6}$Lawrence Livermore National Laboratory, Livermore, CA 94551, USA}\\
\normalsize{\textsuperscript{\textdagger}Current affiliation: Cambridge Heavy Industries, Cambridge, MA 02139, USA}\\
\\
\normalsize{$^\ast$To whom correspondence should be addressed  kanelson@mit.edu}\\
}

\date{}

\begin{document} 


\baselineskip 24pt


\maketitle


\begin{sciabstract}

 Techniques to generate megabar (100 GPa) pressure are critical for the study of materials under extreme conditions. However, kJ-class lasers are typically needed to generate such extreme pressures. Here, we present a method to reach megabar pressure on a benchtop using only tens of mJ a few millijoules of laser pulse energy. The core of our method is the utilization of multiple laser rings of different diameters and time delays to additively excite a focusing shock wave. Using high-speed imaging and a streak camera, we measured the shock trajectory and inferred pressure, temperature and density using hydrocode simulation. Our shock compression efficiency, informed by hydrocode, has achieved 23 GPa/mJ. If scaled up, our method may enable high-energy-density experiments on a benchtop.
 
\end{sciabstract}

\section*{Introduction}
Megabar pressure fundamentally transforms materials by altering electronic, structural, and chemical properties. High pressure can transform materials between insulators and metals, amorphous and crystalline solids, ionic and covalent compounds, and highly reactive and inert substances \cite{mao2018solids}. High-pressure research can be classified into two main approaches: static and dynamic high pressure. Static high-pressure works are commonly conducted using diamond anvil cells (DACs) \cite{hsieh2019imaging} and are widely accessible to individual principal investigator laboratories. The same level of accessibility does not apply to dynamic high-pressure research in the Mbar range which relies mainly on kilojoule laser ablation \cite{millot2018experimental} or flyer plates driven by gas guns \cite{turneaure2017transformation,williams2019structure} in national labs and large-scale user facilities. Those conventional instruments have low repetition rates measured in units of shots per hour or per day, restricting the acquisition of experimental data.

There have been efforts in the miniaturization of high-pressure shock instruments. A benchtop-scale flyer plate driven by a pulsed laser has achieved a few dozens of GPa \cite{dlott2022laser} and a throughput of 100 shots per day. A miniaturized laser-ablation-based shock tube has also been developed for aerodynamic studies \cite{kai2017novel}. A technique for optical generation of cylindrically focusing shocks has shown particular promise in reaching 10s of GPa pressures \cite{pezeril2011direct,Veysset_2015,Veysset_2016,Veysset_2017,Veysset_2018,veysset2019glass}. In that approach, a laser pulse is focused onto the surface of a solid target or into a solid or liquid target that is typically 10 to 50 microns thick. The energy deposition is spatially tailored by shaping the beam into a circular `ring' pattern using an axicon (an axisymmetric conical prism). Depending on the excitation intensity and the nature of the sample, the illuminated ring-shaped region undergoes rapid ablation or melting and evaporation which drives a quasi-2-dimensional converging shock wave. The pressure of the material rises steeply at and near the geometric center point due to conservation of energy in a decreasing volume, as in other examples of focusing shocks \cite{slutz2010pulsed, slutz2012high, atzeni2004physics}. This focusing increases the shock strength far beyond what a planar shock generated by the same laser pulse energy could achieve. The approach also offers continual optical access to the focusing shock which is exploited in the present work for both monitoring and manipulating the shock as it converges. 

Here, we present a method for generating megabar (Mbar) pressures using optically driven two-dimensional converging shock waves. In this approach, multiple concentric laser rings of progressively smaller diameters are launched sequentially, with the largest ring arriving first followed by smaller rings separated by nanosecond-scale delays. The ring diameters and timing are designed to satisfy a velocity-matching condition, allowing each newly generated shock to reinforce the same converging shock front through additive excitation. The combined effects of additive shock reinforcement and geometric focusing substantially amplify the peak pressure. Previous studies have demonstrated nonlinear pressure enhancement through the interaction of two shock waves, including both planar shock addition \cite{lem2026laser} and additive converging shocks \cite{gutierrez2022bullseye} generated using an independent laser cavity for each of two shocks. The present method provides a straightforward and scalable architecture for generating multiple additive focusing shocks from one single laser system.


\section{Experiments}

\begin{figure}[!htb]
\includegraphics[width =1\textwidth]{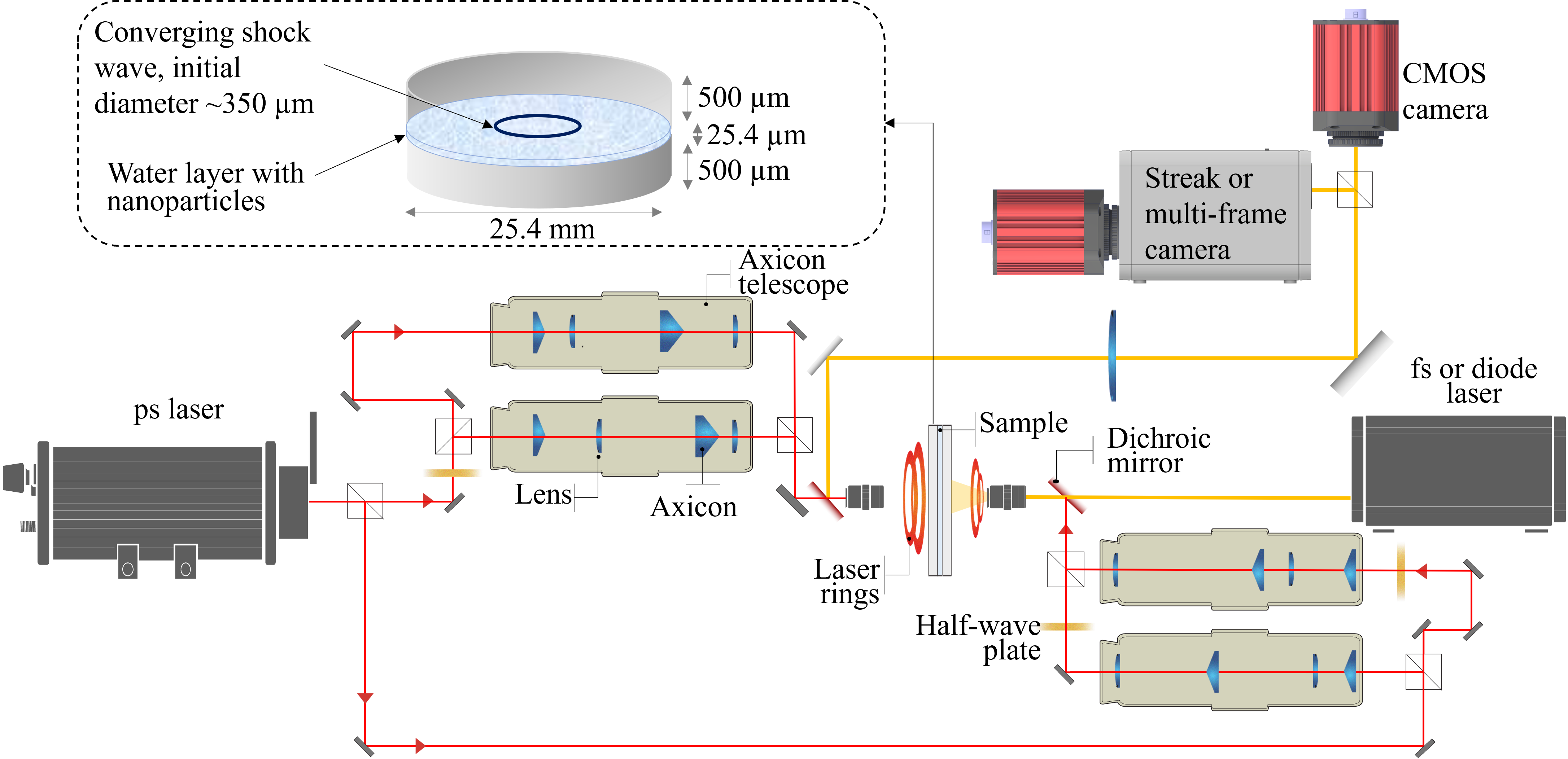}
\caption{The multi-ring method. Multiple axicon telescopes generate multiple laser rings on the sample. A conceptual illustration of the excitation of one focusing shock in the sample is displayed in the dashed box.}
\label{multiRing_setup}	
\end{figure}

Figure \ref{multiRing_setup} depicts the optical configuration used for multi-ring shock excitation. The system is driven by a Nd:YAG laser (Ekspla SL235, 1064 nm, max 100 mJ, 5 Hz, 200 ps pulse duration) that emits pulses with a Gaussian spatial profile. A mechanical shutter permits single-shot operation. Using a series of polarizing beam splitters and wave plates, the output pulse from the laser is split into four temporally delayed pulses. Each pulse propagates through an independent axicon telescope comprising two axicons (axisymmetric conical prisms) and two lenses. The optical path length of each channel is progressively increased to introduce inter-pulse delays of approximately 10 ns.

Each axicon telescope contains precision optomechanics allowing independent alignment of each axicon and lens. Axial translation of the axicons provides micrometer-scale control of the ring diameter, while translation of the lenses determines the position of the focal plane. The x–y positioning of an axicon pair is used to minimize ellipticity and inhomogeneity of the laser ring.

The sample consists of a 25 \textmu m layer of water (defined by a stainless steel spacer), doped with 5 wt\% commercial ink (Epson 502, black, containing carbon nanoparticles), sandwiched between two transparent substrates (25.4 mm diameter, 500 \textmu m thick sapphire unless otherwise specified). An illustration of the sample assembly is displayed in the dashed box of Fig. \ref{multiRing_setup}. A CMOS camera (Hamamatsu Orca) is applied  mainly to observe the sample before and after the shock excitation. The camera is also used to take snapshot images of the shocks in combination with a femtosecond laser (Coherent Vitara, 800 nm, 150 fs). A multi-frame high-speed camera (SIMX16, Specialized Imaging) with a maximum frame rate of 300 million frames per second was applied to capture the laser excitation events. The sample was illuminated with a diode laser (Cavilux, 640 $\pm$ 10 nm wavelength, 30 \textmu s duration). A streak camera (Hamamatsu C13410-01A) is used in combination with a diode laser (Cavilux HF UHS, 640 $\pm$ 10 nm wavelength, 150 ns duration) to record the shock trajectories. Note that the streak camera and multi-frame camera do not work at the same time in our setup, but they are exchangeable.

The throughput of this benchtop setup was about 1 shot per minute, limited mainly by manual rastering of the sample in between shots. Around 1000 shots have been taken to collect useful data. Considering shots for optical alignment, around 5000 shots have been recorded in total.  Automated sample rastering or replacement could reduce the time between shots to a few seconds using commercially available techniques.


\section{Results}

\begin{figure}[!htb]
\centering
\includegraphics[width =0.8\textwidth]{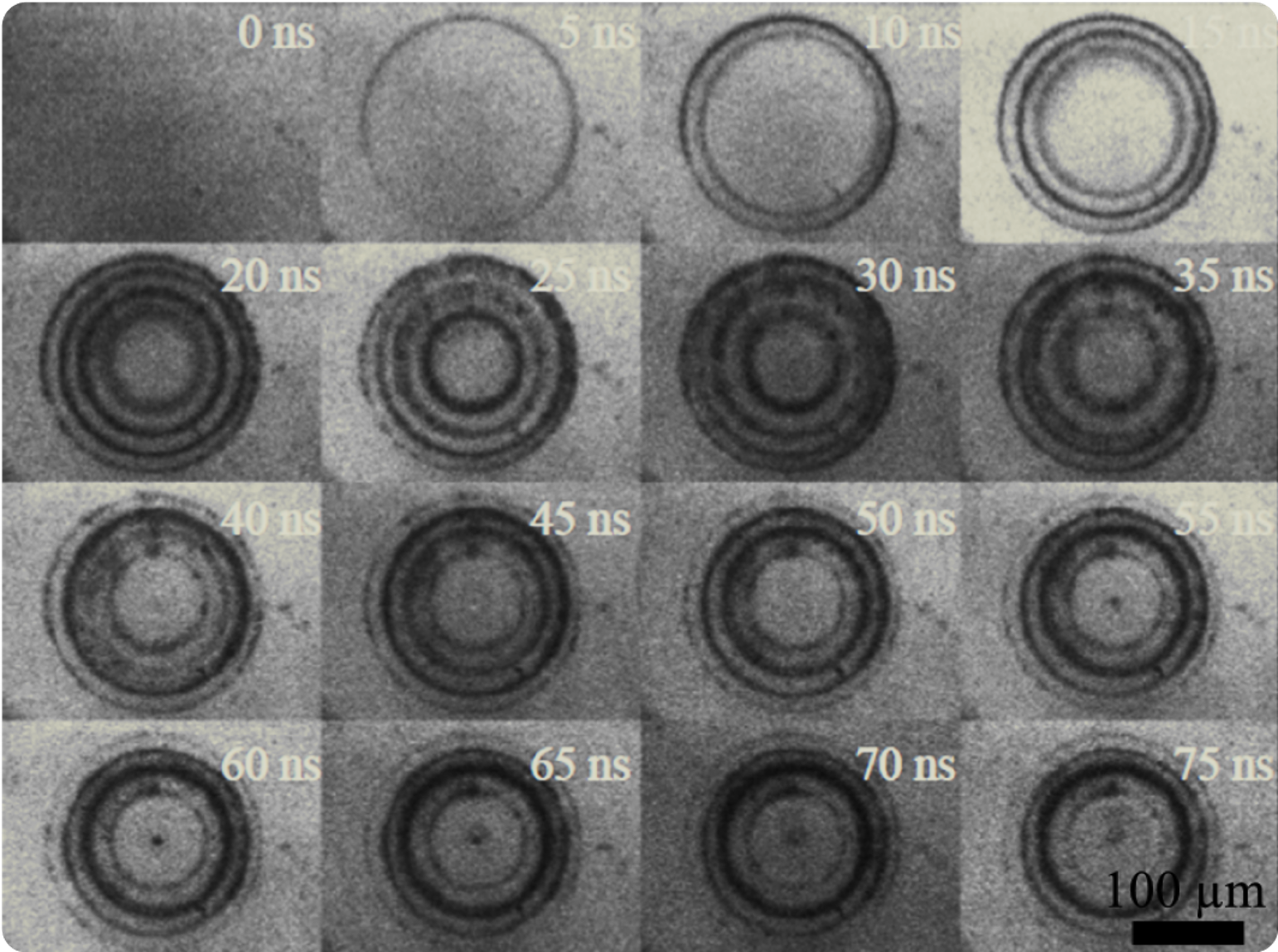}
\caption{Multi-frame high-speed camera images of laser rings additively exciting a water sample. 3.5 mJ incident laser pulse energy per ring. 200 \textmu m thick quartz substrates were applied to confine the water thin film sample.}
\label{HighSpeedCameraImages}	
\end{figure}

High-speed camera images of additive laser ring excitations are shown in Fig. \ref{HighSpeedCameraImages}. The main purpose of this set of data is to register the laser excitation rings and the sample response. The main purpose of this initial set of measurements is to confirm the engineering of multiple laser rings with different diameters and time delays can be delivered to a thin layer of water sample. As we can observe, the multi-ring excitations can be conducted. However, due to the 5 ns time resolution of the high-speed multi-frame camera, shock waves are blurred and not resolvable in this set of images. 



\begin{figure}[!htb]
\includegraphics[width =1\textwidth]{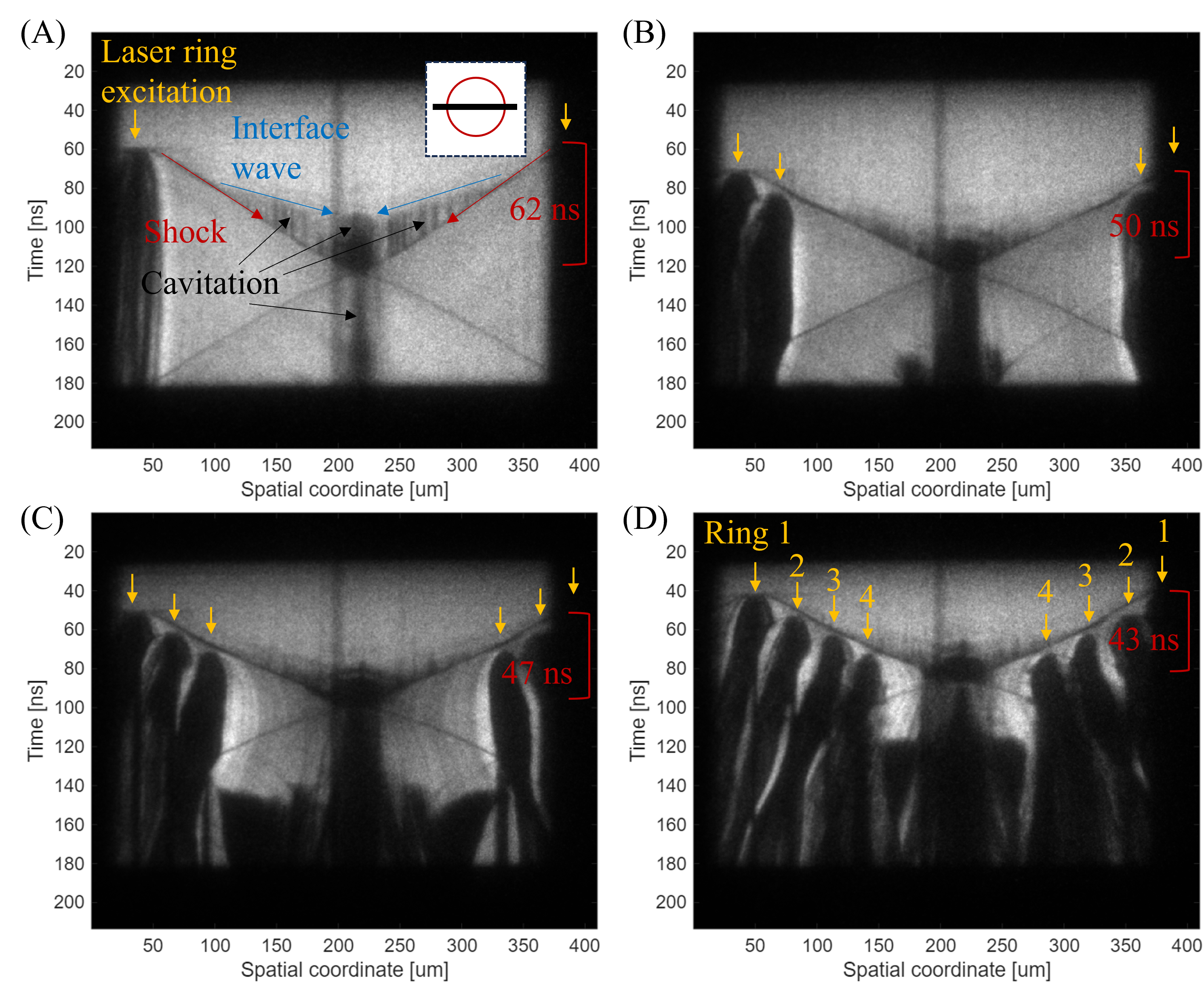}
\caption{Streak camera images of focusing shocks induced by different numbers of laser rings. (A) to (D), 1 ring to 4 rings. The insert (dashed box) indicates the field of view (black solid line) of the streak camera, capturing a quasi 1D slice of the ring. Each laser ring has 4.5 mJ incident energy. Propagation time from the outermost ring to the shock focus is indicated in each sub-figure. Ring radii from the outermost to innermost ring: 169 \textmu m, 134 \textmu m, 105 \textmu m, and 75 \textmu m. Time delay between rings was designed to be 10 ns, nominally; but photodetector measures the actual delays as 11.1 ns between ring 1 and 2; 10.7 ns between ring 2 and 3; 9.2 ns between ring 3 and 4.}
\label{Fig_streak_ringNumVar}	
\end{figure}

To resolve the shock wave, we used a streak camera in setup similar to the works in \cite{pezeril2011direct, lem2026laser}. The streak camera records one spatial coordinate, taken along a line passing through the center of the excitation rings and hence through the shock focus, and one temporal coordinate. These coordinates are displayed along the horizontal and vertical axes, respectively, in Fig. \ref{Fig_streak_ringNumVar}. Figure \ref{Fig_streak_ringNumVar}(A) shows the result for excitation with a single laser ring. The two red arrows indicate the shock fronts propagating inward from the laser-excited region toward its geometric center. The shock reaches the focus after approximately 62 ns and subsequently diverges. The streak image therefore captures the complete trajectory of the converging and subsequently diverging shock wave.

In addition to the shock trajectory, the streak camera also records the evolution of the darkening region. Before the bulk shock wave in water reaches the geometric center, a faster-moving disturbance propagate ahead of the bulk shock, producing an early darkening surrounding the geometric center of the laser ring. This is likely due to an interface wave between the water layer and the sapphire substrate. This interpretation is supported by observation in Figure \ref{Fig_streak_ringNumVar}(A). We can observe that the darkening in the geometric center occurs 30 ns after the laser excitation, which translates to around 5.6 \textmu m/ns wave speed (the propagation distance is 169 \textmu m). The Rayleigh wave speed of sapphire is well known to be around 5.6 \textmu m/ns \cite{aubert2010surface}, which coincides with the measurement. As the bulk shock in water follows up, this darkened region is eliminated (further away from the center) or reduced (at the center). We suspect the darkening may be a result of the interface wave-induced opacity in the sapphire. Substrate deformation and changes in the optical properties of the water are additional possibilities. This darkening feature is an observation with an unresolved physical origin. Its mechanism can be investigated in a follow-up work.


Figures \ref{Fig_streak_ringNumVar}(B) to (D) show the corresponding measurements with additional excitation rings. The additional laser excitation locations coincide with the shock trajectory, achieving a ``speed matching'' condition which leads to additive pumping of the propagating shock. As the number of rings increases, the propagation time from the outermost ring to the shock focus, over a distance of approximately 160 \textmu m, becomes progressively shorter. This reduction in propagation time directly indicates an increase in the shock speed and thus the shock strength.




\section{Hydrocode simulations}

Hydrodynamic simulations using Eul2D were conducted to compare with the experimental results and describe the states achieved along the shock trajectory. Eulerian hydrodynamics was used to accommodate large shear flows, using a multi-material multi-physics hydrocode \cite{Eul2D}. The simulations were run in 2D in the $r-z$ plane, i.e. imposing axial symmetry. $r$ refers the in-plane distance to the shock focus, while $z$ refers to the out-of-plane coordinate. The laser pulse is modeled by heating 10 \textmu m of water uniformly over 200 ps time in the simulation. The water was modeled using the Livermore equation of state (LEOS) model 2017 \cite{LEOS2017}. The sapphire was modeled using LEOS model 2020 \cite{LEOS2020} and elastic-perfectly plastic strength with a shear modulus of 136\,GPa and a flow stress of 20\,GPa \cite{Swift2021}, and negligible tensile strength. These equation-of-state and strength parameters are drawn from internal Lawrence Livermore reports rather than peer-reviewed, publicly documented sources; we note this as a limitation on independent verification of the absolute pressure scale reported below.

\begin{figure}[!htb]
\includegraphics[width =1\textwidth]{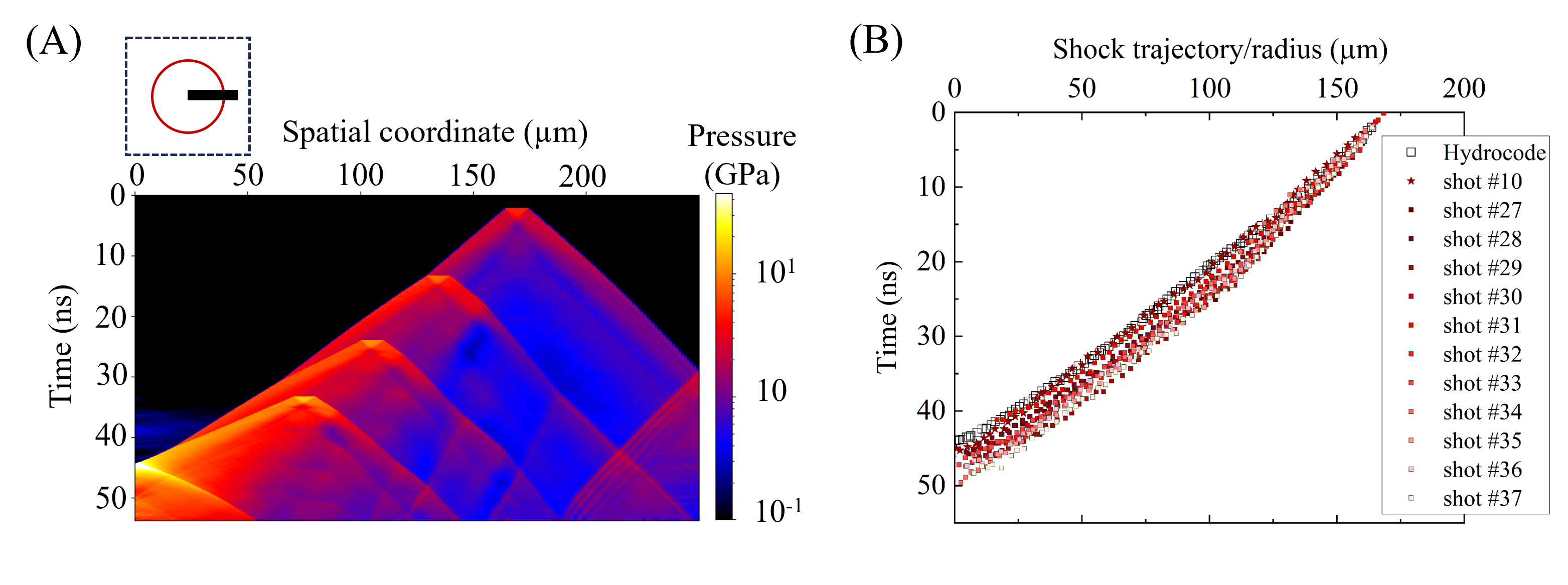}
\caption{Validation of hydrocode. (A) Hydrocode x-t diagram of additive focusing shock induced by four ring excitations, each with 2 mJ energy. The insert (dashed box) indicates the field of view (black solid box) of the simulation, following shock propagation from one location on the laser excitation ring (red) to the focus. (B) Comparison between simulated and measured shock trajectories, i.e. the instantaneous radius of the shock at each time value, where the data symbol size approximately represents the uncertainties in data reading.}
\label{Fig_hydrocode}	
\end{figure}

To validate the hydrocode results, we compare the simulated and measured trajectories in Fig \ref{Fig_hydrocode}. The hydrocode simulation is conducted using the same laser ring diameters and inter-pulse delays as in the experiments. In Fig \ref{Fig_hydrocode}(A), the simulated trajectory is rendered in a streak-camera-like representation to provide visual comparison with the experimental observation, i.e. x-axis is space while y-axis is time. The first excitation is initiated at  $t$ = 2 ns (this time offset is created for clearer illustration, but will be compensated later when compared against experiments) and spatial coordinate $x$ = 169 \textmu m (corresponding to the radius of the first laser ring). The second laser excitation occurs at $t = 13.1$ ns, which is 11.1 ns after the first excitation, at the coordinate $x = 134$ \textmu m. The second laser excitation location coincides with the trajectory of the initial shock, achieving the speed-matching condition. The third ($t = 23.8$ ns, $x = 105$ \textmu m) and fourth ($t = 33.1$ ns, $x = 75$ \textmu m) simulated laser excitations also approximately matched the propagating shock trajectory, resulting in one single combined shock at $x = 20$ \textmu m, before the shock arrived at the focus ($x = 0$). While the incident pulse energy measured after the final optical element (a microscope objective) was 4.5 mJ in the experiment, the effective energy absorbed by the sample and coupled into the shock is expected to be lower. We performed an energy scan in the simulation, from 10 mJ per ring downward to 2 mJ in 1-mJ steps, to investigate the energy deposition. 

By matching the experimentally measured shock trajectories, including the laser-ring velocity-matching conditions, with hydrodynamic simulations, we infer an effective deposited energy of approximately 2 mJ per ring, compared with an incident laser energy of 4.5 mJ per ring. This corresponds to an effective energy-deposition efficiency of approximately 44\%. Independent measurements of laser absorption (Supplementary Fig. S1) indicate absorption efficiencies on the order of 70\% under related experimental conditions. The difference is physically plausible, as absorbed laser energy is initially converted into internal energy of the sample and subsequently redistributed through hydrodynamic motion, shock compression, and heating of surrounding materials. We emphasize that the inferred 44\% represents an effective energy-deposition efficiency calibrated against the measured shock trajectories, rather than a direct measurement of the energy carried by the propagating shock.

We note that this energy-deposition efficiency is obtained by tuning a single free parameter (absorbed energy per ring); while this constrains the hydrocode’s dynamics more strongly than a single travel-time comparison would, it remains a single-parameter calibration and does not by itself provide an independent, direct measurement of the absolute peak-pressure scale reported below. In Fig \ref{Fig_hydrocode}(B), we compare the simulated and measured shock trajectories. The simulated trajectory matched best with the strongest experimental shot, namely shot number 10. We choose the results from the ``best shot'' for further analysis. In our experiments, the laser intensity fluctuation (measured to be around 10\%), excitation inhomogeneity, and instabilities could result in deviations in shock trajectories.

\begin{figure}[!htb]
\includegraphics[width =1\textwidth]{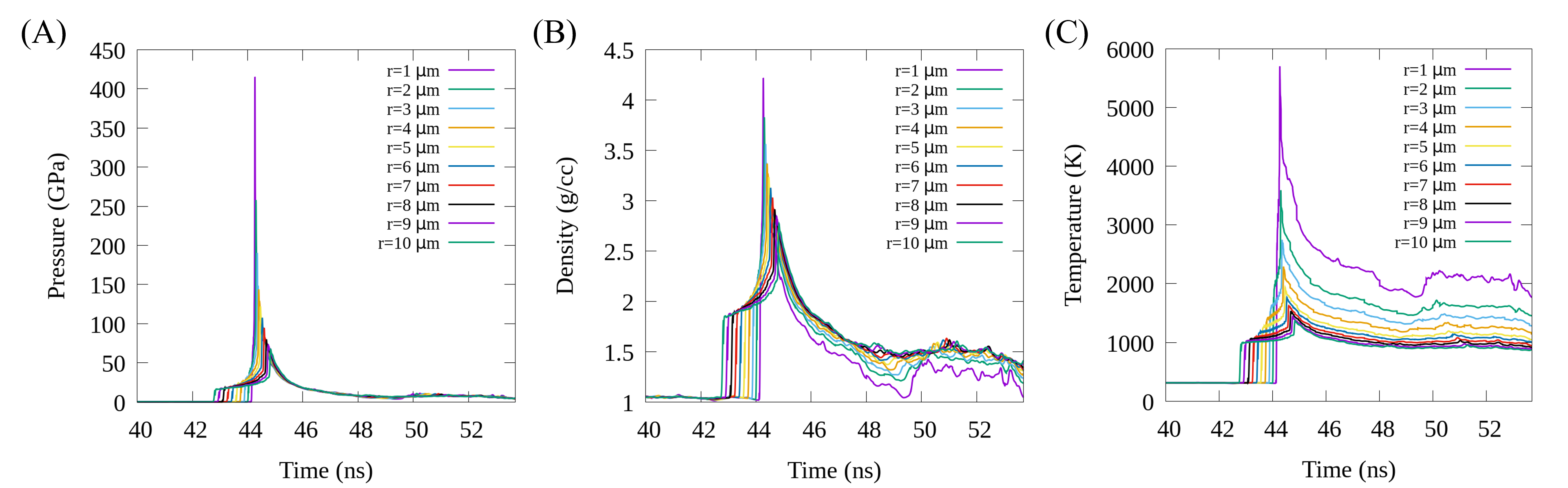}
\caption{Material states extracted from the hydrocode. (A), (B), and (C) show the shock pressure, density, and temperature respectively at different spatial coordinates as a function of time. The simulated additive focusing shock is induced by four laser rings, each with 2 mJ laser pulse energy.}
\label{hydrocode_states}	
\end{figure}

After the validation, we then move on to extract the pressure/density/temperature histories of the shock at different distances from the shock focus. The extracted data from hydrocode calculations are presented in Fig. \ref{hydrocode_states}. The closest location that is still resolvable is 1 \textmu m from the focus, where the shock pressure exceeds 400\,GPa. The corresponding density is 4 g/cm$^3$ and the temperature is about 5700\,K, i.e. $\sim$0.5 eV.

 The simulation results clearly indicate that our multi-shock method can generate Mbar-level pressures at the shock focus using only tens of mJ of incident laser energy. The simulations do not account for deviations from ideal axisymmetry that could be caused by non-uniformity in the excitation laser intensity around each ring and by sample heterogeneity. The deviations would lead to an effective focal spot size that would be given not by the sub-micron shock front but by an averaging across slightly different arrival times of the shock components that converge from different angles. We do not have a quantitative determination of the spot size, but in 2D images that we record, we do not see discernible deviations from circular shock profiles. Thus we conservatively estimate that our effective shock focal spot size must be less than 5 \textmu m. Encouragingly, the simulations indicate that Mbar pressures should be reached even 5 \textmu m from the focus, i.e. within a 10-\textmu m diameter shock focus. On this basis we estimate that we are reaching the Mbar shock pressure range at present, though this conclusion currently relies on hydrocode inference rather than a direct experimental pressure diagnostic. As an outlook, we are seeking quantitative measurements at the focus, for example through the use of x-ray measurements of sample density.

\begin{figure}[!htb]
\includegraphics[width =1\textwidth]{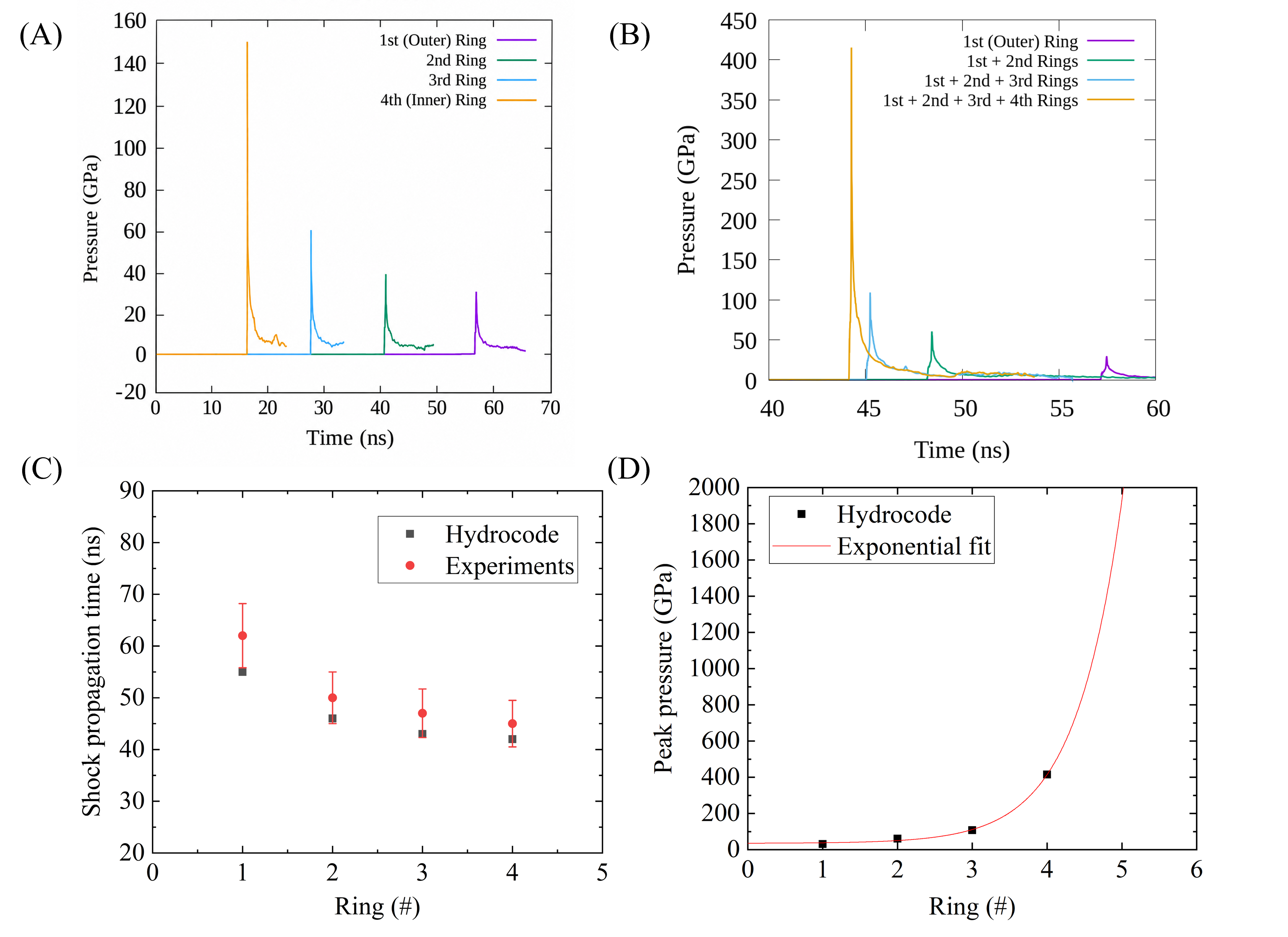}
\caption{Additive effects. (A), simulated pressure histories at the shock focus induced by individual laser rings (without additive pumping); (B), simulated pressure histories at the shock focus corresponding to additively induced focusing shocks. (C), shock propagation time $t$ as a function of the number of laser rings $N$; Error bars represent standard deviations across repeated measurements. (D), shock pressure at the focus as a function of the number of laser rings. 2 mJ is set as the absorbed laser energy per ring in the simulations. In (C), 4.5 mJ is the incident laser energy per ring in experiments.}
\label{Hydrocode_additive}	
\end{figure}

To gain further insight into the additive mechanism responsible for shock enhancement, we examine the hydrocode results presented in Fig.~\ref{Hydrocode_additive}. Figure~\ref{Hydrocode_additive}(A) shows the simulated pressure histories at the shock focus corresponding to individual focusing shocks generated by each laser ring separately. Because the laser pulse energy is held constant at 2 mJ per ring while the ring diameter decreases from the outermost to the innermost ring, the smaller rings achieve higher laser intensities and consequently generate higher peak pressures upon shock convergence. Among the individual rings, the innermost (4th) ring produces the highest peak pressure, reaching approximately 150 GPa at the shock focus. In our current experimental settings, the saturation limit apparently has not been reached, otherwise the simulated shock trajectory would not have agreed with the experimental one. This figure also suggests that the even distribution of laser energy into multiple laser rings may not be the optimal parameter setting, because a outer laser ring could use more laser energy to obtain the same laser intensity as the inner laser ring. 

Figure \ref{Hydrocode_additive}(B) shows the pressure histories at the shock focus for additive excitation using different numbers of laser rings. As the number of rings increases, the peak pressure increases. For four-ring excitation, the peak pressure reaches approximately 420 GPa, substantially exceeding the value expected from linear superposition of the individual-ring contributions. Note that a simple linear summation of the peak pressures generated by the four individual rings yields a total of only 280 GPa. In (C), the additive effects are demonstrated from the aspect of shock propagation time. Stronger shocks correspond to higher shock velocities, shortening the propagation time from the first ring location to the shock focus.

Figure \ref{Hydrocode_additive}(D) shows that the peak pressure exhibits a strongly nonlinear dependence on the number of excitation rings. Over the limited range investigated here, the hydrocode results are well described by an exponential fit: $P = a\cdot\exp(N/b)+c$ where $P$ is the peak pressure at the shock focus and $N$ is the number of rings. This exponential form should be regarded as a phenomenological description rather than a universal scaling law. Physically, the rapid growth can be rationalized by a multiplicative-gain picture: each velocity-matched ring deposits additional energy into an already converging shock, while geometric focusing and nonlinear shock interaction further amplify the shock strength.

Nonlinear shock amplification has previously been reported \cite{lem2026laser}, consistent with the behavior observed here. The experiments were conducted in a planar geometry with a sample assembly similar to ours, but without geometric convergence (using a pair of cylindrically focused, parallel ``line'' rather than ring patterns of pump light) and were limited to pressures in the single-digit gigapascal range. In contrast, the multi-ring method combines additive pumping with energy focusing, allowing successive laser-driven shocks to deposit energy into an already converging compression wave. The simultaneous action of additive shock pumping and geometric energy concentration produces substantially greater pressure amplification and provides a scalable route to extreme pressures.


\section*{Summary and outlook}

We have presented a benchtop platform for generating megabar-level shock pressures using only tens of millijoules of laser energy. The multi-ring method combines additive shock excitation with geometric energy focusing, enabling efficient amplification of converging shocks through the coordinated action of multiple temporally delayed laser rings. By continually depositing energy into an already propagating converging shock, the multi-ring method achieves substantially higher pressures than can be generated by an individual laser ring alone.

The scalability of the multi-ring method offers a pathway toward stronger shocks through additional excitation stages and increased driver energy. Furthermore, the diameters of the laser rings can be readily adjusted so that the laser rings mismatch the initial shock wave, generating a sequence of weaker compression shocks instead of one single combined shock. Such a configuration may enable quasi-isentropic compression, allowing density to increase while minimizing entropy production. This flexibility offered by the multi-ring method will enable a range of applications in materials under extreme conditions.

\section*{Acknowledgment}

We gratefully acknowledge David Veysset, Jet Lem, Calvin Smith, and Scott Wilks for helpful discussions. This research was partially supported by the U.S. Army Research Office under Cooperative Agreements No. W911NF-22-2-0170 and No. W911NF-18-2-0048, and by the Basic Research Office, Office of Under Secretary of Defense for Research and Engineering (OUSD R\&E) under the Laboratory University Collaboration Initiative (LUCI) program. In addition, the MIT Deshpande Center for Technological Innovation has provided support.

\newpage

\nocite{Benson1992} 
\bibliography{references}

@book{williams2019structure,
  title={Structure-property relationships under extreme dynamic environments: shock recovery experiments},
  author={Williams, Cyril L},
  year={2019},
  publisher={Morgan \& Claypool Publishers}
}

@article{millot2018experimental,
  title={Experimental evidence for superionic water ice using shock compression},
  author={Millot, Marius and Hamel, Sebastien and Rygg, J Ryan and Celliers, Peter M and Collins, Gilbert W and Coppari, Federica and Fratanduono, Dayne E and Jeanloz, Raymond and Swift, Damian C and Eggert, Jon H},
  journal={Nature Physics},
  volume={14},
  number={3},
  pages={297--302},
  year={2018},
  publisher={Nature Publishing Group}
}

@article{kai2017novel,
  title={A novel shock tube with a laser--plasma driver},
  author={Kai, Yun and Garen, Walter and Schlegel, Theodor and Teubner, Ulrich},
  journal={Laser and Particle Beams},
  volume={35},
  number={4},
  pages={610--618},
  year={2017},
  publisher={Cambridge University Press}
}

@article{pezeril2011direct,
  title={Direct visualization of laser-driven focusing shock waves},
  author={Pezeril, Thomas and Saini, Gagan and Veysset, David and Kooi, Steve and Fidkowski, Piotr and Radovitzky, Raul and Nelson, Keith A},
  journal={Physical Review Letters},
  volume={106},
  number={21},
  pages={214503},
  year={2011},
  publisher={APS}
}

@article{mao2018solids,
  title={Solids, liquids, and gases under high pressure},
  author={Mao, Ho-Kwang and Chen, Xiao-Jia and Ding, Yang and Li, Bing and Wang, Lin},
  journal={Reviews of Modern Physics},
  volume={90},
  number={1},
  pages={015007},
  year={2018},
  publisher={APS}
}

@article{hsieh2019imaging,
  title={Imaging stress and magnetism at high pressures using a nanoscale quantum sensor},
  author={Hsieh, S and Bhattacharyya, Prabudhya and Zu, C and Mittiga, T and Smart, TJ and Machado, F and Kobrin, B and H{\"o}hn, TO and Rui, NZ and Kamrani, Mehdi and Chatterjee, S and Choi, S and Zaletel, M and Struzhkin, VV and Moore, JE and Levitas, VI and Jeanloz, R and Yao, NY},
  journal={Science},
  volume={366},
  number={6471},
  pages={1349--1354},
  year={2019},
  publisher={American Association for the Advancement of Science}
}

@article{dlott2022laser,
  title={Laser pulses into bullets: tabletop shock experiments},
  author={Dlott, Dana D},
  journal={Physical Chemistry Chemical Physics},
  volume={24},
  number={18},
  pages={10653--10666},
  year={2022},
  publisher={Royal Society of Chemistry}
}

@article{turneaure2017transformation,
  title={Transformation of shock-compressed graphite to hexagonal diamond in nanoseconds},
  author={Turneaure, Stefan J and Sharma, Surinder M and Volz, Travis J and Winey, JM and Gupta, Yogendra M},
  journal={Science Advances},
  volume={3},
  number={10},
  pages={eaao3561},
  year={2017},
  publisher={American Association for the Advancement of Science}
}

@article{gutierrez2022bullseye,
  title={Bullseye focusing of cylindrical waves at a liquid--solid interface},
  author={Guti{\'e}rrez-Hern{\'a}ndez, Ulisses J and Reese, Hendrik and Ohl, Claus-Dieter and Quinto-Su, Pedro A},
  journal={Physics of Fluids},
  volume={34},
  number={11},
  year={2022},
  publisher={AIP Publishing}
}

@article{veysset2019glass,
  title={Glass fracture by focusing of laser-generated nanosecond surface acoustic waves},
  author={Veysset, David and Kooi, Steven E and Haferssas, Ryadh and Hassani-Gangaraj, Mostafa and Islam, Mohammad and Maznev, AA and Chernukha, Yevheniia and Zhao, Xiaoguang and Nakagawa, Keiichi and Martynowych, Dmitro and Zhang, Xin and Lomonosov, Alexey M and Schuh, Christopher A and Radovitzky, Raul and Pezeril, Thomas and Nelson, Keith A},
  journal={Scripta Materialia},
  volume={158},
  pages={42--45},
  year={2019},
  publisher={Elsevier}
}

@article{Veysset_2015,
author = {Veysset,D.  and Pezeril,T.  and Kooi,S.  and Bulou,A.  and Nelson,Keith A. },
title = {Laser-induced versus shock wave induced transformation of highly ordered pyrolytic graphite},
journal = {Applied Physics Letters},
volume = {106},
number = {16},
pages = {161902},
year = {2015}
}

@article{Veysset_2016,
author = {Veysset,D.  and Maznev, Alexei A. and  Pezeril, T. and  Kooi,S. and  Nelson, Keith A.},
title = {Interferometric analysis of laser-driven cylindrically focusing shock waves in a thin liquid layer},
journal = {Scientific Reports},
volume = {6},
number = {16},
year = {2016}
}

@article{Veysset_2017,
author = {Veysset,David  and Maznev,A. A.  and Veres,István A.  and Pezeril,Thomas  and Kooi,Steven E.  and Lomonosov,Alexey M.  and Nelson,Keith A. },
title = {Acoustical breakdown of materials by focusing of laser-generated Rayleigh surface waves},
journal = {Applied Physics Letters},
volume = {111},
number = {3},
pages = {031901},
year = {2017}
}

@article{Veysset_2018,
  title = {Single-bubble and multibubble cavitation in water triggered by laser-driven focusing shock waves},
  author = {Veysset, D. and Guti\'errez-Hern\'andez, U. and Dresselhaus-Cooper, L. and De Colle, F. and Kooi, S. and Nelson, K. A. and Quinto-Su, P. A. and Pezeril, T.},
  journal = {Physical Review E},
  volume = {97},
  issue = {5},
  pages = {053112},
  numpages = {8},
  year = {2018},
  month = {May}
}

@article{lem2026laser,
  title={Laser shocking a shock wave for nonlinear summation of GPa pressures},
  author={Lem, Jet and Kai, Yun and Vassaux, Maxime and Kooi, Steven E and Nelson, Keith A and Pezeril, Thomas},
  journal={Physical Review E},
  volume={113},
  number={1},
  pages={015502},
  year={2026},
  publisher={APS}
}

@misc{Eul2D,
  author       = {D. C. Swift},
  howpublished = {unpublished notes on {\sc Eul2D} hydrocode (physics and numerics in ref.~\cite{Benson1992}), Los Alamos National Laboratory},
  year         = {2003}
}

@article{Benson1992,
  author  = {D. J. Benson},
  title   = {Computational methods in Lagrangian and Eulerian hydrocodes},
  journal = {Computer Methods in Applied Mechanics and Engineering},
  volume  = {99},
  pages   = {235--394},
  year    = {1992}
}

@misc{LEOS2017,
  author       = {P. A. Sterne},
  institution  = {Lawrence Livermore National Laboratory},
  howpublished = {unpublished notes on {{\sc LEOS} 2017} equation of state for water, Lawrence Livermore National Laboratory},
  year         = {2022}
}

@misc{LEOS2020,
  author       = {D. A. Young},
  institution  = {Lawrence Livermore National Laboratory},
  howpublished = {unpublished notes on {{\sc LEOS} 2020} equation of state for sapphire, Lawrence Livermore National Laboratory},
  year         = {2001}
}

@misc{Swift2021,
  author       = {D. C. Swift},
  institution  = {Lawrence Livermore National Laboratory},
  howpublished = {unpublished notes on a strength model for sapphire, Lawrence Livermore National Laboratory},
  year         = {2021}
}

@book{atzeni2004physics,
  title={The Physics of Inertial Fusion: Beam Plasma Interaction, Hydrodynamics, Hot Dense Matter},
  author={Atzeni, Stefano and Meyer-ter-Vehn, J{\"u}rgen},
  number={125},
  year={2004},
  publisher={Oxford University Press}
}

@article{slutz2010pulsed,
  title={Pulsed-power-driven cylindrical liner implosions of laser preheated fuel magnetized with an axial field},
  author={Slutz, SA and Herrmann, MC and Vesey, RA and Sefkow, AB and Sinars, DB and Rovang, DC and Peterson, KJ and Cuneo, ME},
  journal={Physics of Plasmas},
  volume={17},
  number={5},
  year={2010},
  publisher={AIP Publishing}
}

@article{slutz2012high,
  title={High-gain magnetized inertial fusion},
  author={Slutz, Stephen A and Vesey, Roger A},
  journal={Physical Review Letters},
  volume={108},
  number={2},
  pages={025003},
  year={2012},
  publisher={APS}
}

@article{aubert2010surface,
  title={Surface acoustic wave devices based on AlN/sapphire structure for high temperature applications},
  author={Aubert, Thierry and Elmazria, Omar and Assouar, Badreddine and Bouvot, Laurent and Oudich, Mourad},
  journal={Applied Physics Letters},
  volume={96},
  number={20},
  year={2010},
  publisher={AIP Publishing}
}

\bibliographystyle{Science}

\clearpage

\end{document}